# Shear-mode Direct Piezoelectric Response of Ferroelectric Nematic Liquid Crystals

Péter Salamon[1], Marcell Tibor Máthé[1], Hiroya Nishikawa[2], Fumito Araoka[2], Antal Jákli[1,3,4]

[1]Institute for Solid State Physics and Optics, HUN-REN Wigner Research Centre for Physics, P.O. Box 49, Budapest H-1525, Hungary

[2]RIKEN Center for Emergent Matter Science, 2-1 Hirosawa, Wako, Saitama, 351-0198, Japan

[3]Department of Physics, Kent State University, Kent, Ohio 44242, USA

[4]Materials Sciences Graduate Program and Advanced Materials and Liquid Crystal Institute, Kent State University, Kent, Ohio 44242, USA

Corresponding authors: Péter Salamon:salamon.peter@wigner.hu; Antal Jákli: ajakli@kent.edu

**Abstract:**

Piezoelectricity (linear coupling between mechanical deformation and electric signal) was originally observed only in solid crystals. Recently, it was discovered that liquid ferroelectric nematic liquid crystals are also piezoelectric. However, so far only their converse piezoelectric signals (applied voltage-induced mechanical deformation) were measured quantitatively.

In this work, we have carried out periodic shear-induced electric current and oscillatory rheology measurements on the two archetypic ferroelectric nematic compounds, RM734 and DIO. From temperature, frequency and strain dependent results of the first and second harmonic current signals together with the results from oscillatory rheometry, we were able to quantitatively determine the shear-mode direct piezoelectric coupling constants. These values are similar for both materials and are compared to results of previous converse piezoelectric measurements. We propose a physical mechanism in which the flow alignment of ferroelectric polarization leads to the direct piezoelectric response.

## I. Introduction

Ferroelectric materials are polar, therefore lack inversion symmetry. As such they are also piezoelectric, i.e., possess linear coupling between mechanical and electric fields (linear electromechanical effects). Ferroelectricity was originally known only in solid crystals, but later it was found in smectic and columnar liquid crystals with one- and two-dimensional spatial periodicity[1] and recently, even in nematic liquid crystals with no spatial periodicity but only 3-dimensional fluid order.[2–8] In spite of their genuine 3D fluidity, in electric fields they often behave as solid particles or microrobots[9], they can also form metastable freestanding filaments[10,11] that can be stabilized in small electric fields[11].

Similar to ferroelectricity, piezoelectricity was first observed only in solid crystals, then in polymers. From the late 70s linear electromechanical effects were also observed in ferroelectric smectic phases of chiral rod-shaped[12,13] and achiral bent-core molecules[14], furthermore in the columnar phases of chiral disc-shaped molecules.[15] Very recently, soon after their discovery in 2017, electromechanical effects were also reported in ferroelectric nematic liquid crystals by Máthé et al.[16] and Rupnik et al.[17]

The mathematical definition of the direct piezoelectricity, $\Delta P_i = \sum_{jk} \gamma_{ijk} T_{jk}$, dictates that the elements of the stress tensor $T_{jk}$ that induces a polarization component $\Delta P_i$ proportional to the elements $\gamma_{ijk}$ of the third rank piezoelectric charge coupling tensor, has to be effective on fluids that flow under DC stress. Similarly, the definition of the converse piezoelectricity, $S_{jk} = \sum_i \gamma_{ijk} E_i$, requires that the strain $S_{jk}$ induced by an external electric field applied along $i$, should be small and reversible. Both requirements mean that in fluids we can speak about piezoelectricity only when either the stress in direct effect or the electric field in the converse effect are periodic and small enough to lead to small strain $\hat{S} = \vec{\nabla} \cdot \vec{s}$. Here the displacement of a volume element is $\vec{s}(t) = \vec{s}_o \cdot e^{i\omega t} \ll L$, where $\omega$ is the angular frequency of the vibration, and $L$ is film thickness.[18]

So far, quantitatively only some of the converse piezoelectric responses were measured. It was found that in room temperature ferroelectric nematic liquid crystal films aligned parallel to the substrates along axis 3 an electric field applied between the substrates (along 1 in Figure 1), there is a periodic vibration along the polarization corresponding to $\gamma_{1,13} \sim 3 \frac{nm}{V} = 3\ nC/N$ below $f \sim 1\ kHz$ and at higher frequencies the piezoelectric coupling constant is inversely proportional to the frequency.[18] These values are larger than that of the strongest piezoelectric crystals and comparable to that of soft ferroelectrets[19–22] due to the softness of the ferroelectric nematic fluids with loss moduli $G" = \eta \cdot \omega$, where $\eta \sim 0.1 - 1\, Pa \cdot s$ is a typical flow viscosity of room temperature FNLC materials. This angular frequency dependence of the loss modulus explains the observed $\gamma_{1,13} \propto \omega^{-1}$ for the $N_F$ fluids.[18] Interestingly, converse piezoelectric response was also measured normal to the substrates (direction 3 in Figure 1) along the applied electric field with frequency dependences consisting of a set of resonant peaks in the kHz frequency range[16,18]. This response is not expected by symmetry and likely related to imperfect alignment, a pretilt or the fact that the top plate is not exactly parallel to the bottom one, therefore a motion in 1 direction is coupled to a vertical motion (direction 3).

Direct piezoelectric-type response was observed by Rupnik et al.[17] using a simple demonstrator device, where the ferroelectric nematic liquid was placed into a deformable container with electrodes, and the electric current induced by both periodic and irregular actuation of the container was examined. They found that $100\ \mathrm{Hz}$ actuation frequency and $100\ \mathrm{k}\Omega$ load resistance may generate $\sim 10\ \mu\mathrm{W}/\mathrm{cm}^2$ power per unit area which is only 0.01% of the theoretically estimated current and to the specific power produced by solar photovoltaic panels at natural illumination conditions.

In this paper, we use low frequency shear stress to measure direct piezoelectric signals of the two prototypical ferroelectric nematic liquid crystal materials RM734 and DIO and obtain the first quantitative values of the shear piezoelectric charge coupling constant. The values obtained are two orders of magnitude higher than the ones obtained from converse piezoelectric measurements of other ferroelectric nematic liquid crystals. With the help of simultaneous optical observations, we propose a physical model and estimate the maximum shear-induced electric current one may achieve in uniformly poled samples. Furthermore, we present oscillatory rheological properties of RM734 and DIO, where we found soft elastic behavior in the antiferroelectric intermediate phase.

## II. Materials and Methods

In our experiments we used the two archetypic ferroelectric nematic compounds: 2,3′,4′,5′-tetrafluoro[1,1′-biphenyl]-4-yl 2,6-difluoro-4-(5-propyl-1,3-dioxan-2-yl) benzoate (DIO)[3], and 4-[(4-nitrophenoxy)carbonyl] phenyl-2,4-dimethoxybenzoate (RM734[2] from Instec, USA). The reported phase sequences on cooling of DIO and RM734 are I 158 °C N 84.5 °C $SmZ_A$ 68.8 °C $N_F$ and Iso – 189 °C –N – 133°C –$N_F$ ~89 °C, respectively. Here $SmZ_A$ is an antiferroelectric phase[23], also called as splay nematic[24] ($N_s$).

All measurements were carried out using an Anton Paar MCR502 torsional rheometer in oscillatory mode (Figure 1). The lower plate was fixed, the upper plate was oscillated with the torsion angle $\phi(t) = \Delta\phi \cdot \sin(2\pi f t)$, where $\Delta\phi$ is the amplitude of the angle oscillation, $f$ is the frequency and $t$ is the time. The rheometer is operated in parallel plate geometry with gap $h = 80\ \mu m$ between plates and diameter $D = 25\ mm$ of the disc-shape plates. The strain $S$ at the circumference is $S_{D/2} = \frac{D}{2}\frac{\Delta\phi}{h}$. This means that $\Delta\phi = 1\ mrad$, corresponds to $S_{D/2} = 0.15625 = 15.625\ \%$. Sketches of the relevant parameters in the shear geometry and the proposed shear alignment leading to $\gamma_{1,13}$ are shown in Figure 1(a).

For optical observations, we performed experiments with RM734 in the ferroelectric nematic phase at 125°C in between transparent indium-tin-oxide (ITO) coated parallel plates during oscillatory shear using the rheo-optical module of the rheometer with custom modifications (see Figure 1(b)). The diameter of upper rotating plate was 40 $mm$, and the middle of the field of view was at $r = 17.7$ mm radius and, similar to all other experiments, the gap was $h = 80$ µm. The periodically sheared sample between crossed polarizers was illuminated with a filtered monochromatic LED light source at $\lambda = 660$ nm wavelength. The vertical axis of the images was placed at $\pm 45°$with respect to the polarizers and approximately tangential to the direction of motion of the upper plate. Images were taken by a FLIR BFS-U3-28S5M-C camera as a function of phase $\alpha = 2\pi f t$ of the strain. Note that images (see an example in the inset of Figure 1(b)) corresponding to increasing phases were taken in subsequent periods at increasing delays of the camera trigger with respect to the actual strain.

Illustration of the torsional rheometer for measuring phase dependences of current signals. The graphs of the phase dependence of the current generated in DIO at 68 °C and $f$ = 10 Hz at low strain ($\Delta\phi$ = 0.7 mrad) with mainly first harmonic response and with dominant second harmonic at high strain ($\Delta\phi$ = 10 mrad), are shown in Figure 1(c).

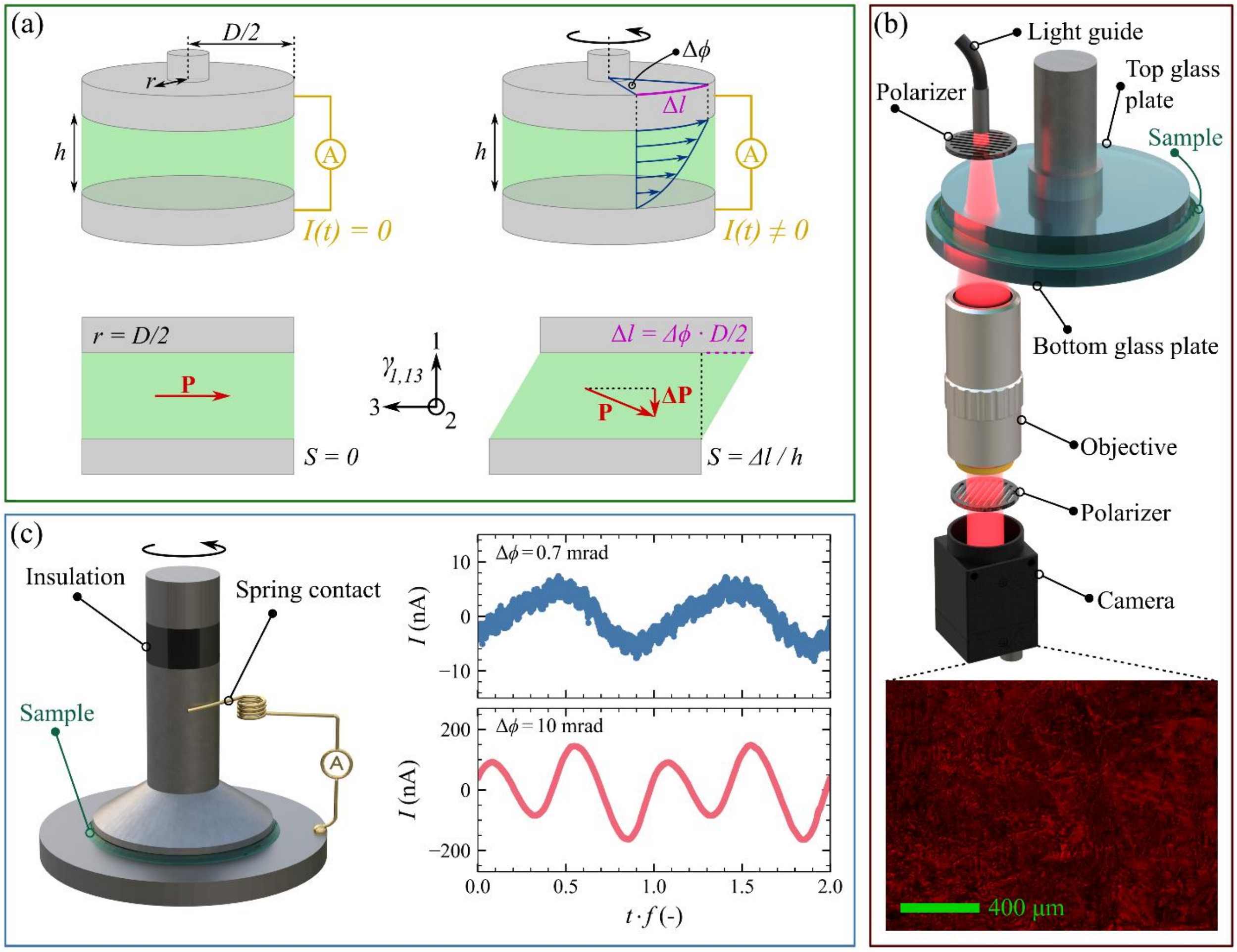

*Figure 1: Sketch and images of parts of the experimental setup and typical measured current and micrograph of the texture. (a) Sketches of the relevant quantities in the shearing geometry and the proposed shear alignment leading to $\gamma_{1,13}$. (b) Illustration of the modified rheometer allowing polarizing optical observations in transmission mode, and a typical micrograph of the sample. (c)Illustration of the torsional rheometer for measuring phase dependences of current signals. The graphs in the inset show current signals from DIO at 68 °C and f = 10 Hz at low strain ($\Delta\phi$ = 0.7 mrad) with mainly first harmonic response and with dominant second harmonic at high strain ($\Delta\phi$ = 10 mrad).*

A single filling of the rheometer with $D = 25\,mm$ diameter measuring system required about $120\,mg$ of material, which was used almost for a week. The samples were kept in the rheometer under $180\,L/h$ continuously heated $N_2$ stream to avoid degradation and maintain the desired temperature. Except for the optical measurements, the rheometer was equipped with the dielectro-rheological module and a PP25/DI/TI measuring system allowing floating electrical connections to the bottom and top sides of the sheared fluid. The electric connection at the moving upper plate was achieved by a gold sliding spring contact (see Figure 1(b)).

The shear induced electrical current between the lower and upper plates was measured by a low-noise current preamplifier (SR570, Stanford Research Systems, USA). The same $c_I = 50\, nA/V$ current conversion factor was used for all measurements. The time dependence of the oscillation angle and the induced current were monitored simultaneously in channel 1 and channel 2 of an oscilloscope (TiePie Handyscope HS5, The Netherlands). The first and second harmonic signals were calculated by digital signal processing using lock-in technique, as $I_{1f}(t) = I_{1f} \cdot \sin(2\pi f t + \psi_{1f})$ and $I_{2f}(t) = I_{2f} \cdot \sin(4\pi f t + \psi_{2f})$, where $I_{1f}$ and $I_{2f}$ are the amplitudes and $\psi_{1f}$ and $\psi_{2f}$ are the phase of the induced current, respectively. Typical unprocessed current signals of DIO at 68 °C and $f$ = 10 Hz can be seen in Figure 1(c).

## III. Results

Temperature dependences of periodic shear induced currents and the phases for RM734 and DIO measured under 0.5 °C/min cooling rate at frequency of $f = 10\ Hz$ and oscillation amplitude of $\Delta\phi = 1$ mrad are shown in Figure 2(a) and (b), respectively.

In the nematic (N) phase of both materials the induced current is very small ($< 0.2$ nA). In spite of it, the fluctuation of the phase that decreases at $< 5\ °/°C$ rate is small indicating that the signal is greater than the measurement error. The induced current increases considerably upon transition to the $N_F$ phase of RM734 and to the $SmZ_A$ and $N_F$ of DIO. For both materials, the temperature dependences of the current have maxima in 1-2 °C range below the phase transitions. Specifically, for RM734 the current rises to about $10\ nA$ at $\approx 1.5\ °C$ below the N-$N_F$ transition, then decreases to about $0.6\ nA$ by 2.2 °C below the transition. Simultaneously the phase drops by about 90° indicating a considerable change in the dynamical response of the material. On cooling below this range, the phase increases by ~180°. Simultaneously, the current increases up to about $\approx 6$ nA at $\approx 127$ °C, then slowly decreases to about $\approx 1.7$ nA by $120\ °C$. We note that the $2.2$ °C temperature range below the N phase is close to the range where recent calorimetric studies showed the presence of an intermediate phase[25] which was shown to depend on the ionic content and is assigned to either an antiferroelectric smectic Z ($SmZ_A$) phase[26] or to a double splayed nematic phase[27].

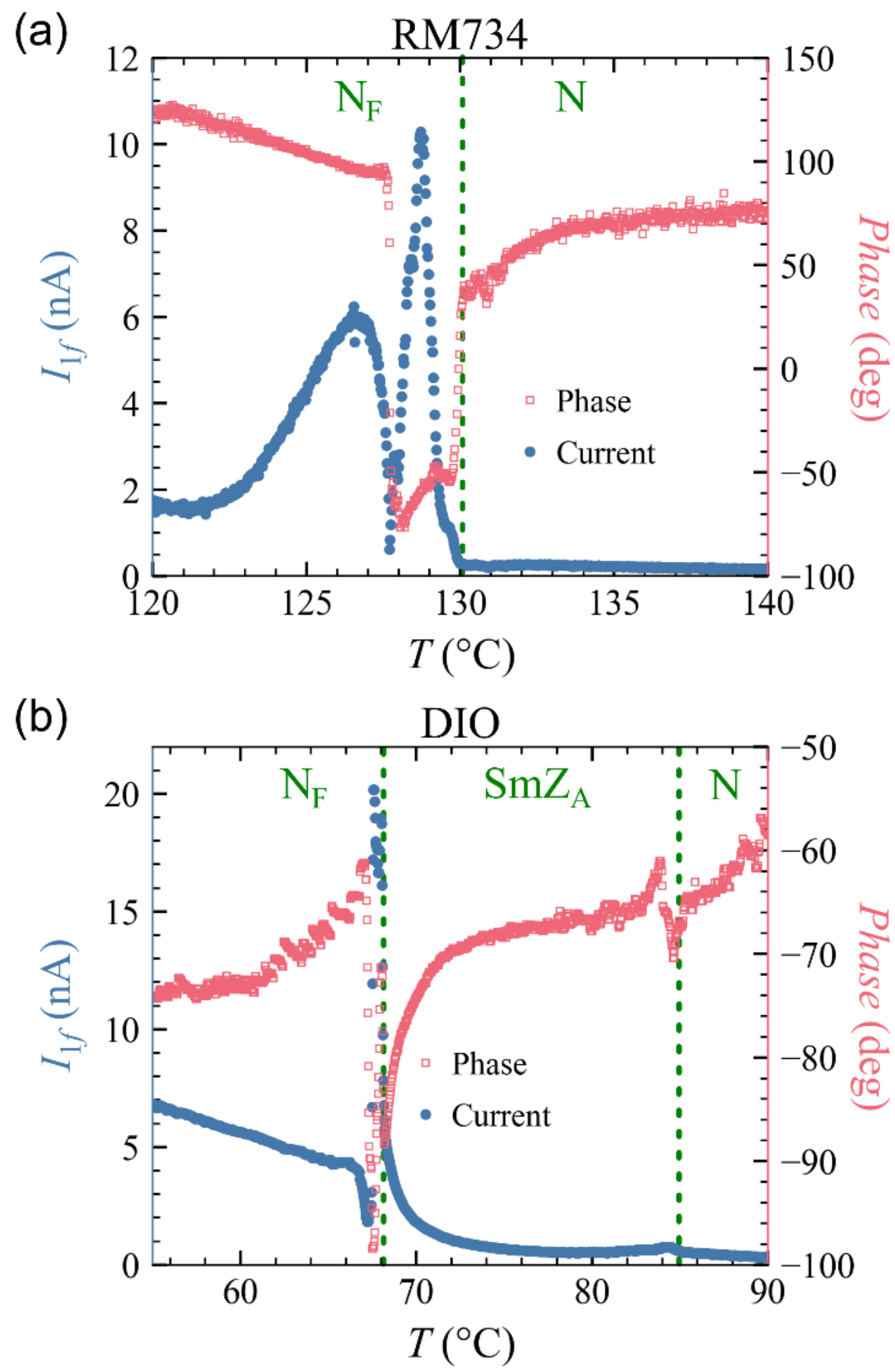

*Figure 2: Temperature dependences of the $I_{1f}$ signals (amplitude and phase) induced by $f = 10\ Hz$, $\Delta\phi = 1\ mrad$ periodic shear for RM734 (a) and DIO (b).*

In DIO, the peak currents reach 0.8 nA, and 20 nA upon the N-SmZ$_A$ and SmZ$_A$- N$_F$ transitions, respectively. Simultaneously, in both transitions there is a transient decrease of the phase. In contrast to RM734, on cooling further in the N$_F$ phase, the amplitude of the shear-induced piezo current increases.

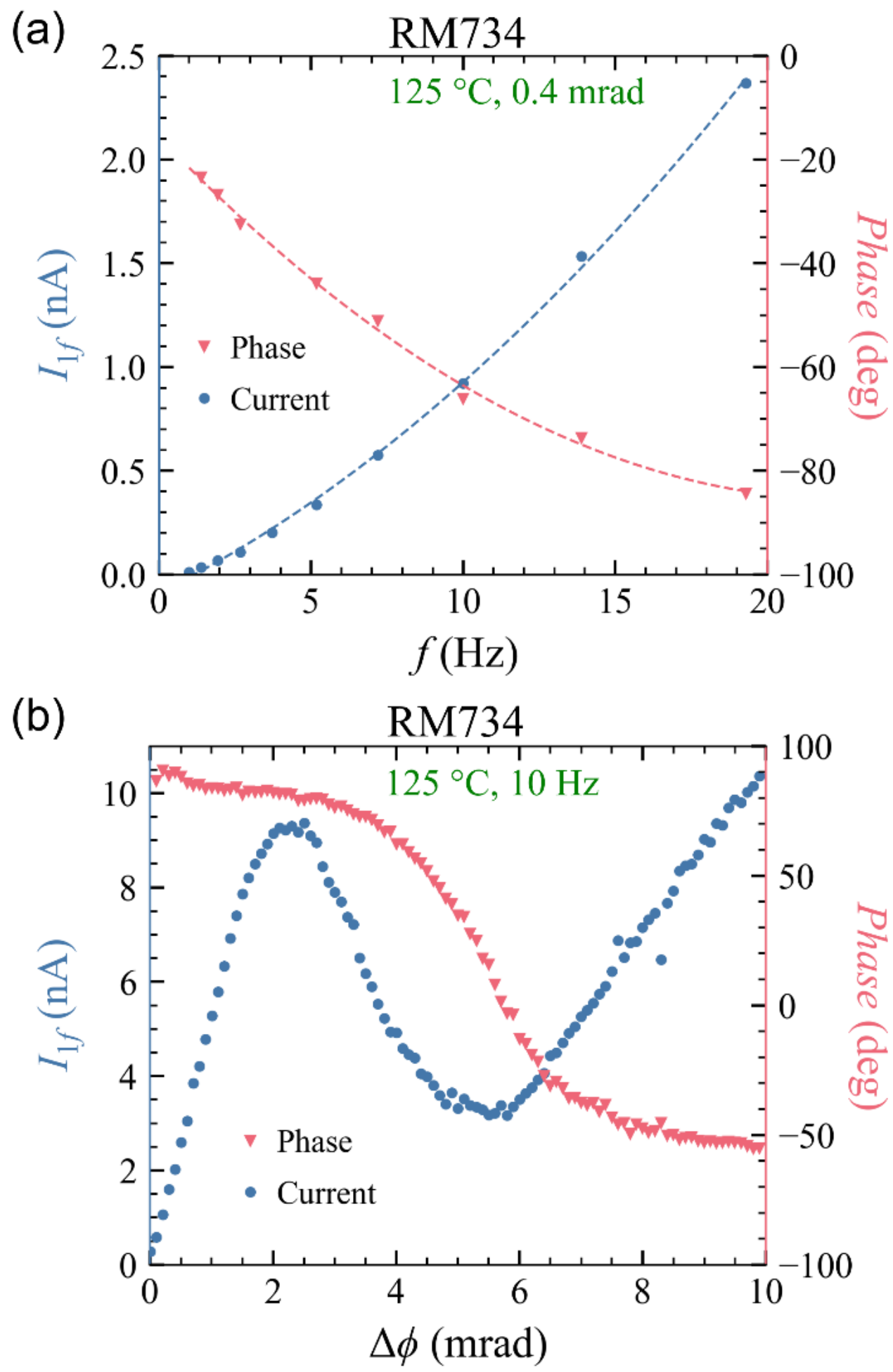

*Figure 3: Frequency and the oscillation angle dependences of the induced first harmonic current in RM734 in the $N_F$ phase at 125 °C. (a) Frequency dependence at $\Delta\phi = 0.4\, mrad$; (b) $\Delta\phi$ dependence at $f = 10\, Hz$.*

The frequency $(f)$ and the oscillation angle $(\Delta\phi)$ dependences of the first harmonic current amplitude $(I_{1f})$ in RM734 in the $N_F$ phase at $125\ °C$ are shown in Figure 3(a) and (b), respectively.

The frequency dependence of current up to $f = 20\ Hz$ at $\Delta\phi = 0.4$ mrad is about linear between 5 Hz and 20 Hz (corresponding to $2\ s^{-1} \leq \dot{S}_{D/2} = \frac{\pi f D \Delta\phi}{h} \leq 8\ s^{-1}$ shear rate). The phase decreases smoothly by about 90° which may indicate variation of the mechanical response likely due to flow alignment. The oscillation amplitude dependence of the current at $f = 10$ Hz up to about $\Delta\phi \approx 2$ mrad (corresponding to $\dot{S}_{D/2} \leq 4\ s^{-1}$ shear-rate) is linear. In the 2 mrad $\leq \Delta\phi \leq$ 6 mrad range the induced current is decreasing and the phase shifts by about 90°. For $\Delta\phi \geq$ 6 mrad the current increasing linearly again but the phase makes about 180° with respect to the low shear rate values, indicating opposite direction of the motion.

The $f$ and $\Delta\phi$ dependences of the induced current in DIO are shown at various temperatures between 55 °C and 90 °C (in the $N_F$, $SmZ_A$ and N phases) in Figure 4(a) and (b), respectively. In accordance with the temperature dependence seen in Figure 2(b), the highest signal is measured at 68 °C upon the $SmZ_A$-$N_F$ transition, and in the vertical axis scale (up to 3 nA) the signals at all other temperatures look very small (within the range marked by the green dotted rectangle). For this reason, the range of the green dotted rectangle (up to 0.3 nA) is blown up in the inset showing that, except in the N phase, the current is proportional to the frequency, i.e., to the shear rate. We see that the data is well fitted by quadratic functions (solid lines in Figure 4(a)).

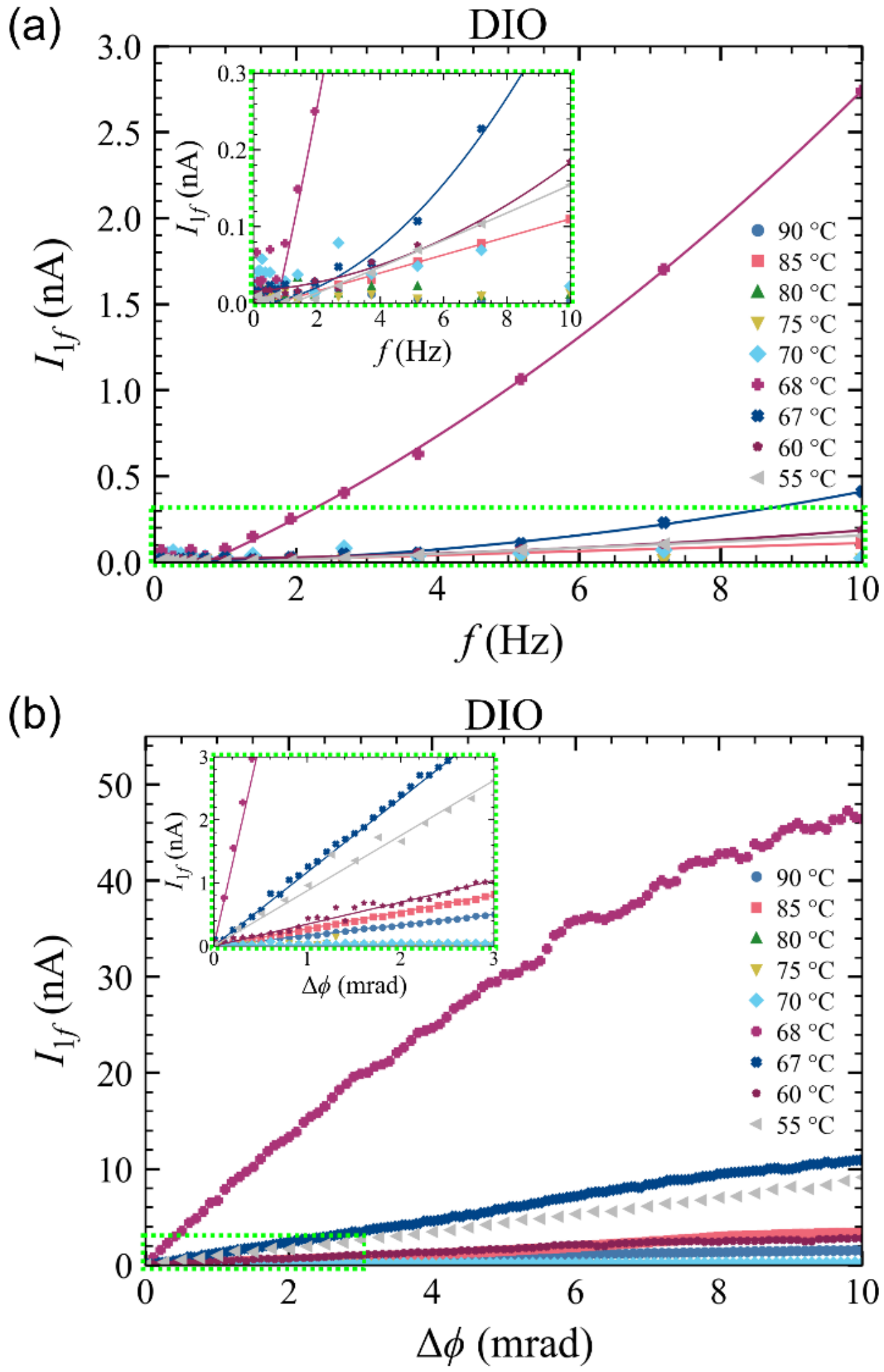

*Figure 4: Frequency and oscillation angle dependences of the induced current in DIO at various temperatures between 55°C and 90 °C. (a) Frequency dependences at $\Delta\phi = 0.4\,mrad$; (b) $\Delta\phi$ dependences at $f = 10\,Hz$.*

Similarly to the frequency dependence, the $\Delta\phi$ dependence is also the largest at 68 °C with a linear increase (Figure 4(b)). The inset shows the response up to $\Delta\phi = 3$ mrad and up to 3 nA range. At all temperatures the currents are proportional to the oscillation angle up to about $\Delta\phi = 3$ mrad, although the slopes are very small in the $N$ phase

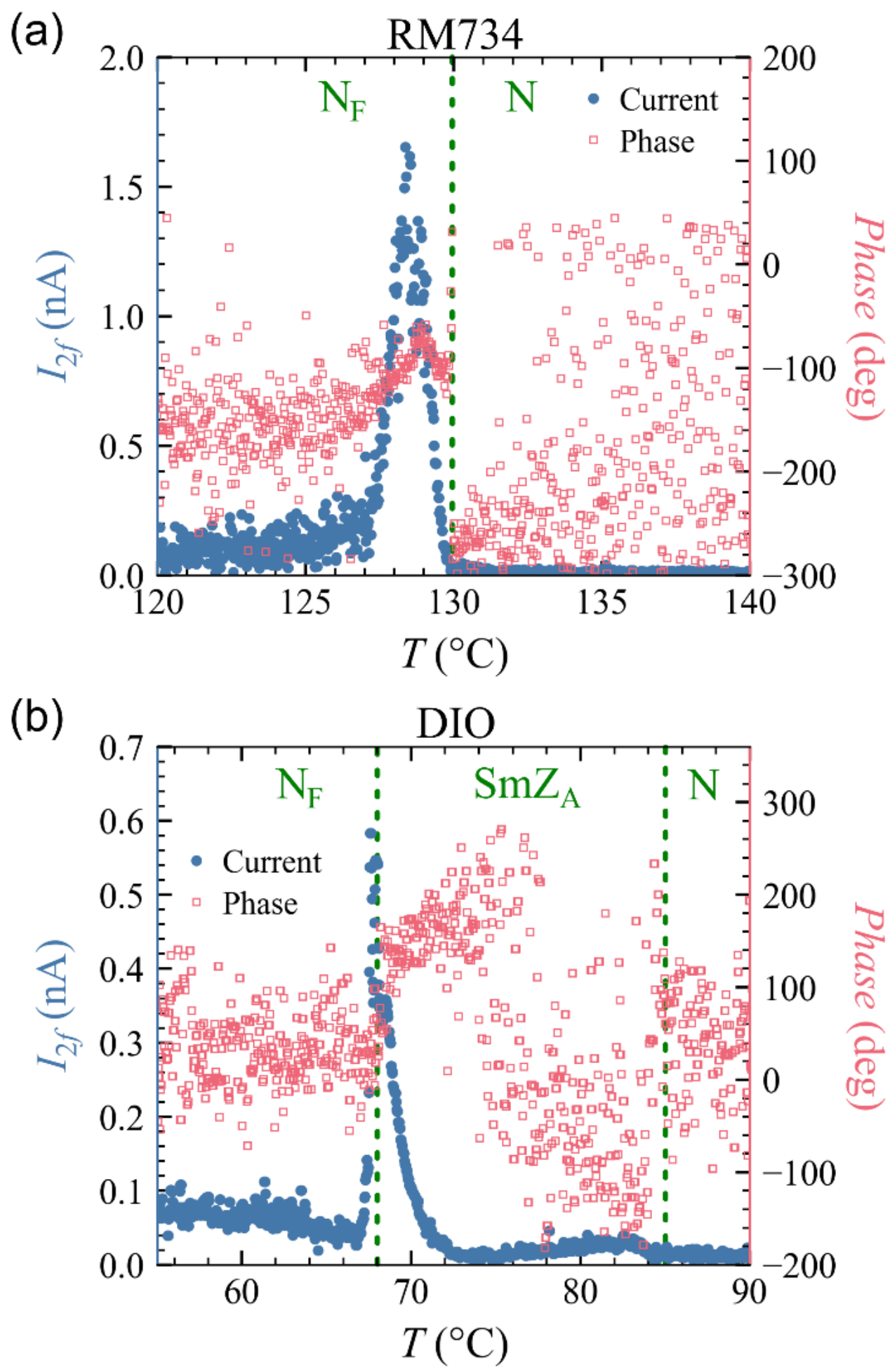

*Figure 5: Temperature dependences of the $I_{2f}$ signals and phases induced by $f = 10\,Hz$, $\Delta\phi = 1\,mrad$ periodic shear for RM734 (a) and DIO (b).*

The temperature dependences of the second harmonic amplitude $I_{2f}$ induced by $f = 10$ Hz, $\Delta\phi = 1$ mrad periodic shear are shown for RM734 and DIO in Figure 5(a) and (b), respectively. They were measured simultaneously at the same conditions as of the 1st harmonic amplitude $I_{1f}$ seen in Figure 2 and show about an order of magnitude smaller signals than of $I_{1f}$. Due to this small signal, the phase data are very noisy, especially in the N phase. Just like in the case of the first harmonic signals, the largest responses are measured at the transition to the $N_F$ phase. Additionally, in case of DIO, there is a smaller maximum at the N-$SmZ_A$ phase transition.

From the $\Delta\phi$ dependences of $I_{1f}$ and $I_{2f}$ of RM734 and DIO at 10 Hz shown in Figure 6, we see that the second harmonic signals overcome the first harmonic signals at oscillation angles $\Delta\phi > 4\,mrad$.

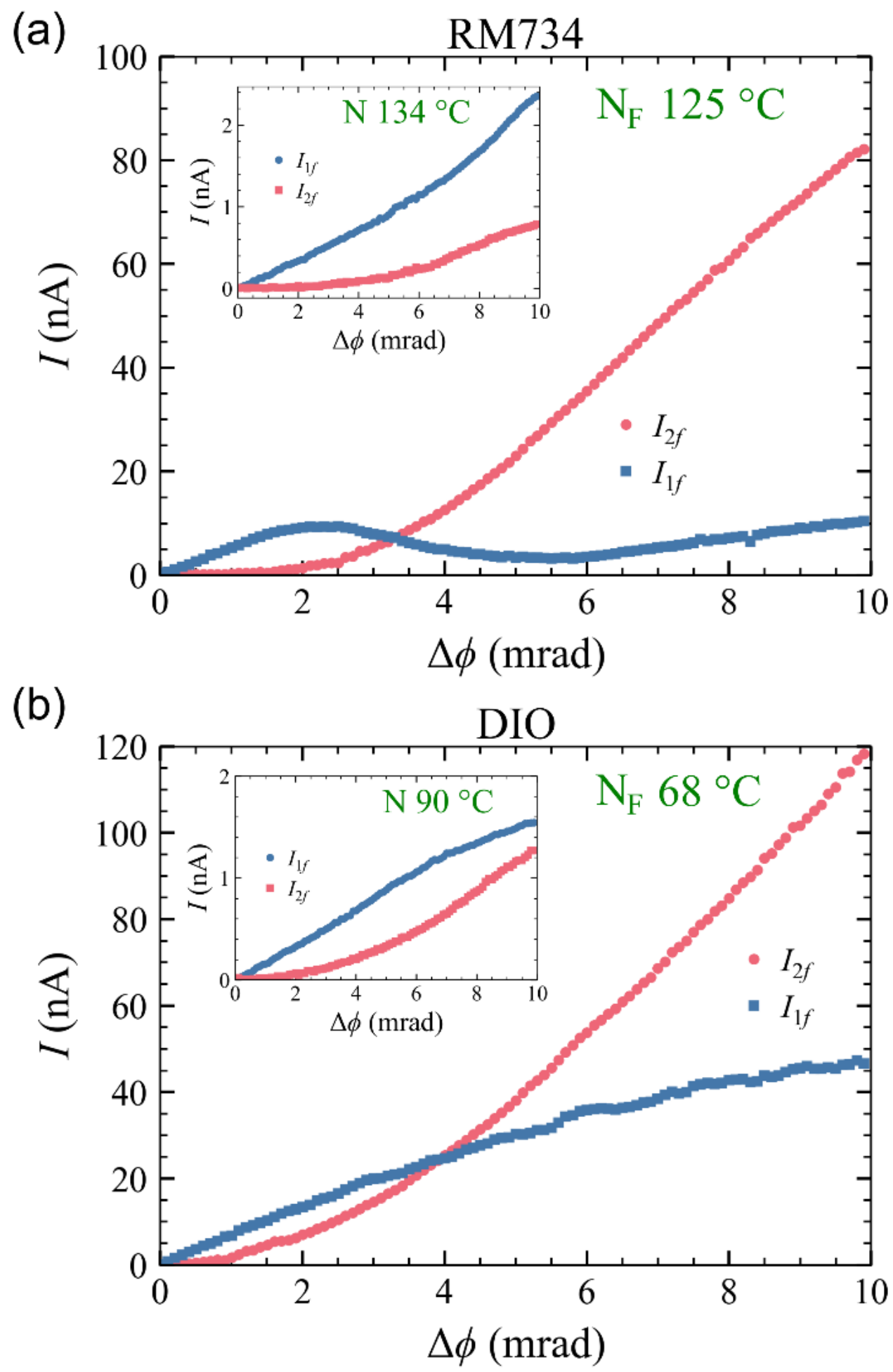

*Figure 6: Oscillation angle dependences of* $I_{1f}$ *and* $I_{2f}$ *at* 10 *Hz. (a) RM734 at 125 °C in the* $N_F$ *phase (main pane) and at 134 °C in the N phase (inset); (b) DIO at 68 °C in the* $N_F$ *phase (main pane) and at 90 °C in the N phase (inset).*

One can also observe that in the $N_F$ phase at $f = 10$ Hz $\Delta\phi > 4$ mrad ($\dot{S} > 4\ \mathrm{s}^{-1}$) the slope of $I_{1f}$ response decreases and the initially quadratic $\Delta\phi$ dependence of $I_{2f}$ becomes almost linear. In the N phase both the $I_{1f}$ and $I_{2f}$ responses are about 2 orders of magnitudes smaller.

## IV. Discussion

As discussed in a recent review[18], the $N_F$ phase has $C_{\infty v}$ symmetry, i.e., continuous rotational symmetry about the polar axis, and mirror symmetry through any plane containing the polar axis 3. This means that any 180° rotation around the polar axis 3 (a symmetry operation) changes the sign of 2 (or 1), therefore, components of $\gamma_{i,jk}$ containing an odd number of either 1 or 2, must be zero. Thus, the non-vanishing piezoelectric tensor elements are $\gamma_{3,33}, \gamma_{3,11}(=\gamma_{3,22})$ and $\gamma_{1,13}(=\gamma_{2,23}=\gamma_{1,31}=\gamma_{2,32})$. In our experiments the polarization direction (along axis 3) can be assumed to be tangential[28], i.e., along the periodic displacement that leads to a velocity gradient normal to substrates (along axis 1), in the direction where we measure the induced current. Therefore, in our studies we measure the piezoelectric charge constant $\gamma_{1,31}=\gamma_{1,13}$.

Our experimental results on the induced current $I_{1f}$ and the measured viscosity data allow us to determine the relevant piezoelectric coupling constant $\gamma_{1,13}$ using the mathematical definition that $\Delta P_1=\gamma_{1,13}\cdot T_{13}=\gamma_{1,13}\cdot S_{13}\cdot G$, where $T_{13}$ and $S_{13}$ are the relevant stress and strain tensor components, respectively, and G is the shear modulus. As an estimate, we identify $\Delta P_1=\langle\Delta P\rangle$ and introduce the average strain $\langle S\rangle=\frac{2}{3}S_{D/2}=\frac{D}{3}\frac{\Delta\phi}{h}$ commonly used in parallel plate torsional rheometry[29]. Then the first harmonic current amplitude is approximated by $I_{1f}=fD^2\pi^2\langle\Delta P\rangle=fD^2\pi^2\langle S\rangle\,\gamma_{1,13}G=\frac{fD^3\pi^2\Delta\phi|G*|}{3h}\,\gamma_{1,13}$. Using the complex viscosity $|\eta *|=\frac{|G*|}{2\pi f}$ we get

$$\gamma_{1,13}=\frac{3\,I_{1f}h}{2\pi^3f^2D^3\Delta\phi\,|\eta*|} \quad (1)$$

To determine $\gamma_{1,13}$, we measured the temperature dependences of the complex viscosities at $f=10$ Hz, which are shown in Figure 7(a) and Figure 7(b) for RM734 and DIO, respectively.

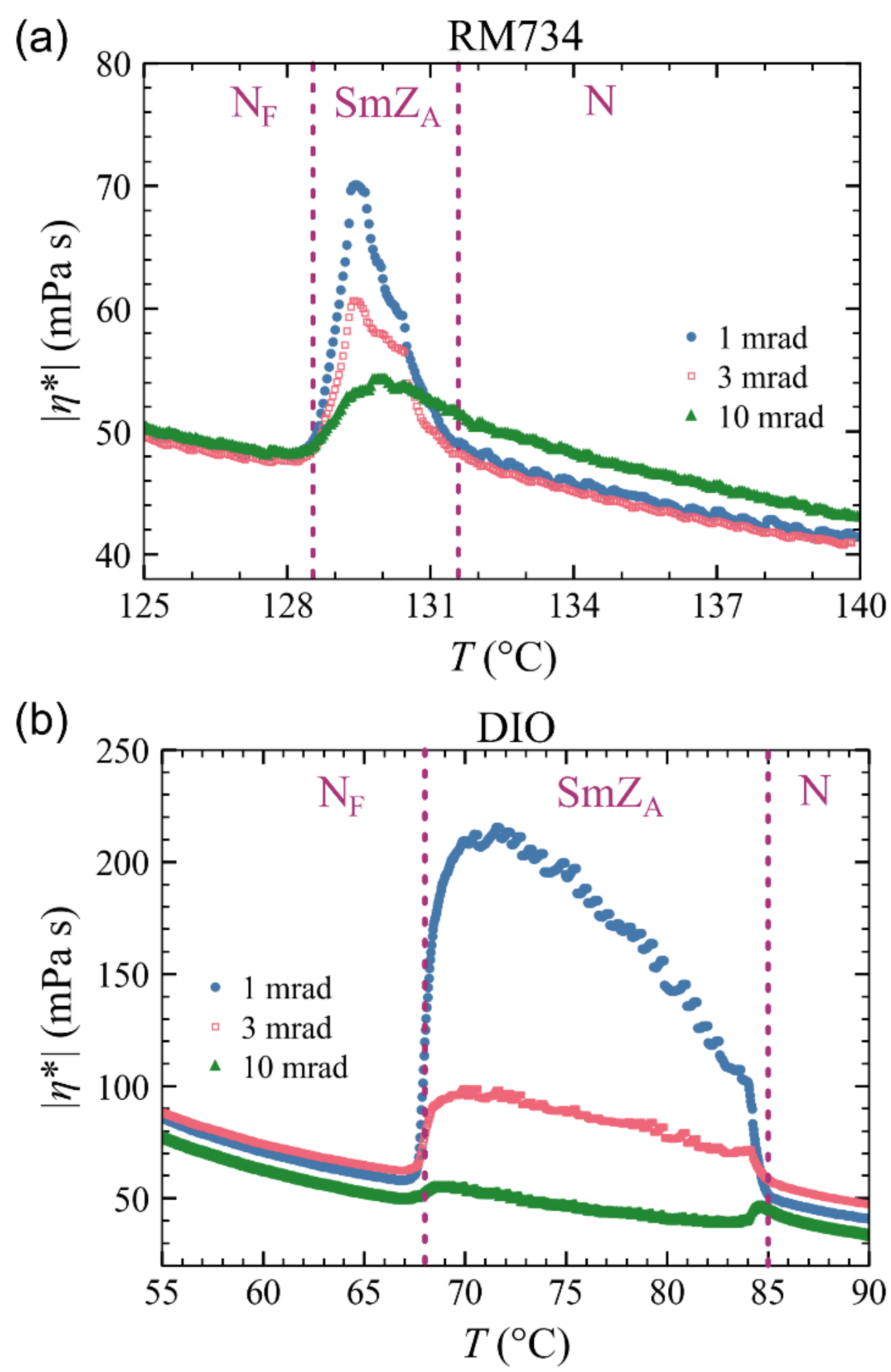

*Figure 7: Temperature dependences of the absolute value of the complex viscosity of RM734 (a) and DIO (b) measured at $f = 10\ Hz$.*

These measurements and (1) give for RM734 at $128\ °C$ in the $N_F$ phase where $I_{1f} = 10\ \text{nA}$ and $\eta = 48\ \text{mPa} \cdot \text{s}$ for $f = 10\ \text{Hz}$, $\Delta\phi = 1\ \text{mrad}$ with $h = 80\ \mu\text{m}$ and $D = 25\ \text{mm}$ that $\gamma_{1,13} \approx 0.5\ \mu\text{C/N}$. For DIO at 68 °C $I_{1f} \approx 20\ \text{nA}$ and $\eta \approx 60\ \text{mPa} \cdot \text{s}$ for $f = 10\ \text{Hz}$, $\Delta\phi = 1\ \text{mrad}$, which gives $\gamma_{1,13} \approx 0.8\ \mu\text{C/N}$. These numbers are 1 and 2 orders of magnitudes larger than the values of $\gamma_{1,13}$ obtained by converse piezoelectric measurements on a room-temperature ferroelectric nematic liquid crystal KPA-02[16] at 100 Hz and 1 kHz respectively. The different values of the piezoelectric coupling constants obtained from converse piezoelectric measurements at $> 100$ Hz and from direct measurements below 10 Hz may be explained by the $\frac{1}{f^2}$ frequency dependence (see in Eq. (1)) of the piezoelectric signals.

The experimental conditions and the proposed physical mechanism leading to $\gamma_{1,13}$ are shown in Figure 1(a). We propose that the physical mechanism leading to the periodic shear-induced electric current is the flow alignment that rotates the polarization toward the electrodes leading to a periodic electric field and a periodic electric current (see Figure 1(a)). The measurable current is the temporal derivative of the charge difference between the top and bottom plates, which can originate from the time dependence of the polarization component along direction 1 at the electrodes, which we denote by $\Delta P(r,t)$ assuming radial dependence. The amplitude of the first harmonic current response is calculated as $I_{1f} = 4\pi f \int_o^{D/2} \int_0^{2\pi} \Delta P(r,t) r dr d\phi = f D^2 \pi^2 \langle \Delta P \rangle$, where we estimate $\langle \Delta P \rangle$ as an average of $\Delta P$. By considering an unrealistic extreme case, when in each half of the oscillation period the spontaneous polarization is entirely oriented along direction 1, the maximum current $I_{1f-max}(f)$ would correspond to $\langle \Delta P \rangle_{max} = P$. This means $I_{1f-max}(f) = f D^2 \pi^2 P \approx f \cdot 6.25 \cdot 10^{-4}\, \pi^2 \cdot 5 \cdot 10^{-2} \approx f \cdot 3 \cdot 10^{-4}$ A, so for $f = 10$ Hz shear, the theoretically achievable maximum induced current is 3 mA. A more realistic assumption is to consider the ferroelectric nematic as a shear aligning material as it was inferred from previous rheological measurements.[30] Other studies indicated that that the shear aligning nature of $N_F$ materials is restricted to shear rates lower than about 40 s$^{-1}$ and at higher shear rates the director aligns along the vorticity direction called the log-rolling regime.[31] In our experiments, the shear rate was significantly below 40 s$^{-1}$, so we stayed in the flow aligning regime. Typically, in conventional shear aligning nematic liquid crystals, the shear alignment angle of the director with respect to the bounding plane is $\Delta\theta \approx$ 5°-15°. Taking $\Delta\theta$ ~6° for our $N_F$ materials, $\langle \Delta P \rangle_{max} = P \sin 6°$, providing $I_{1f-max}(10\ Hz) \approx 0.3$ mA. In our experiments, the highest $I_{1f}$ at $f = 10$ Hz was 90 nA for DIO at 68 °C upon the transition to the $N_F$ phase. This is 3-4 orders of magnitude smaller than the theoretical limit. Assuming completely uniform alignment and rotation of the polarization, this can be interpreted as the induced $\Delta P/P \approx 3 \cdot 10^{-5}$ or the flow alignment angle $\Delta\theta \approx 3 \cdot 10^{-5}$ rad~0.002 degrees. However, the induced current can be low even for much larger director rotation if the director of the polarization were random or even for uniform optical axis by assuming almost equal probability of the polarization pointing left or right. The observation that the double frequency current $I_{2f}$ overcomes the first harmonic signal $I_{1f}$ at $\Delta\phi > 4$ mrad, suggests that $\Delta\theta \gg 0.1$ mrad and simply the oppositely oriented polarizations cancel each other.

To obtain more information about the possible structural changes in the periodically sheared fluid, we recorded the texture of the liquid crystal as a function of the phase, $\alpha = 2\pi f t$, of the strain as seen in Supplementary Video 1 for RM734 at 125 °C with $f$ = 10 Hz and $\Delta\phi =$ 1 mrad. In the studied frequency and strain ranges, no significant intensity change could be detected as a function of α. We can estimate the resulting intensity modulation by assuming flow alignment discussed above. Considering the sheared fluid between crossed polarizers at 45° with respect to the flow velocity, the ξ transmittance ($0 < \xi < 100$ %) can be calculated as $\xi_{\Delta\theta} = \sin^2(\Delta\psi(\Delta\theta)/2)$, where the $\Delta\psi$ optical phase difference is related to the $\Delta\theta$ tilt angle due to flow alignment assuming homogeneous tilt: $\Delta\psi(\Delta\theta) = \frac{2\pi}{\lambda} h \left( n_e \middle/ \sqrt{1 + \frac{n_e^2 - n_o^2}{n_o^2} \sin^2(\Delta\theta)} - n_o \right)$, where $n_e$ and $n_o$ are the extraordinary and ordinary refractive index, respectively. Considering $n_e = 1.6$, $n_o = 1.4$, λ = 660 nm and $h = 80$ μm, the planar initial state leads to $\Delta\psi(0) \approx 48.5\pi$. As being between a minimum ($\Delta\psi = 48\pi$) and a maximum ($\Delta\psi = 49\pi$) intensity range, the transmitted light intensity near the planar initial state is sensitive to the modulation of the tilt angle. In such state, the $\Delta\theta = 3 \cdot 10^{-5}$ rad calculated above results in $\xi_{\Delta\theta} - \xi_0 \approx 8 \cdot 10^{-6}$ %, which can be difficult to detect. Even a hundred times larger tilt angle $\Delta\theta = 3 \cdot 10^{-3}$ rad would result in hardly measurable intensity modulation: $\xi_{\Delta\theta} - \xi_0 \approx 8 \cdot 10^{-2}$ %.

It is however seen that due to the alignment layers, an inhomogeneous alignment occurs with domains of orientation nucleated at the surfaces shown by defect lines. Apparently, the effect of shearing does not result in significant changes in the orientation. Only small alterations may happen, which cannot be resolved by the applied optical detection method, however, this agrees with the above calculation referring to a low level of shear alignment.

In Figure 8, we have plotted the intensity (integrated on the entire field of view) as a function of phase, α, for different Δϕ values for RM734 at 125 °C and $f$ = 10 Hz. The data show undetectable intensity changes at $\Delta\phi < 1$ mrad, while at $\Delta\phi \geq 4$ mrad, a small (~0.1%) modulation is seen that is comparable to $\xi_{\Delta\theta} - \xi_0 \approx 8 \cdot 10^{-2}$ % calculated above. We also note that the appearance of the second harmonic signal is seen at $\Delta\phi = 20$ mrad. This angle is larger than what we observed from the electromechanical measurements, which indicates that the mechanical effect of the second harmonic component of the director is much stronger than its optical effect. It is because the opposite direction of the rotation adds up in the $I_{2f}$ but tends to cancel out in the optical signal as seen in in Supplementary Video 2, for $\Delta\phi = 20$ mrad.

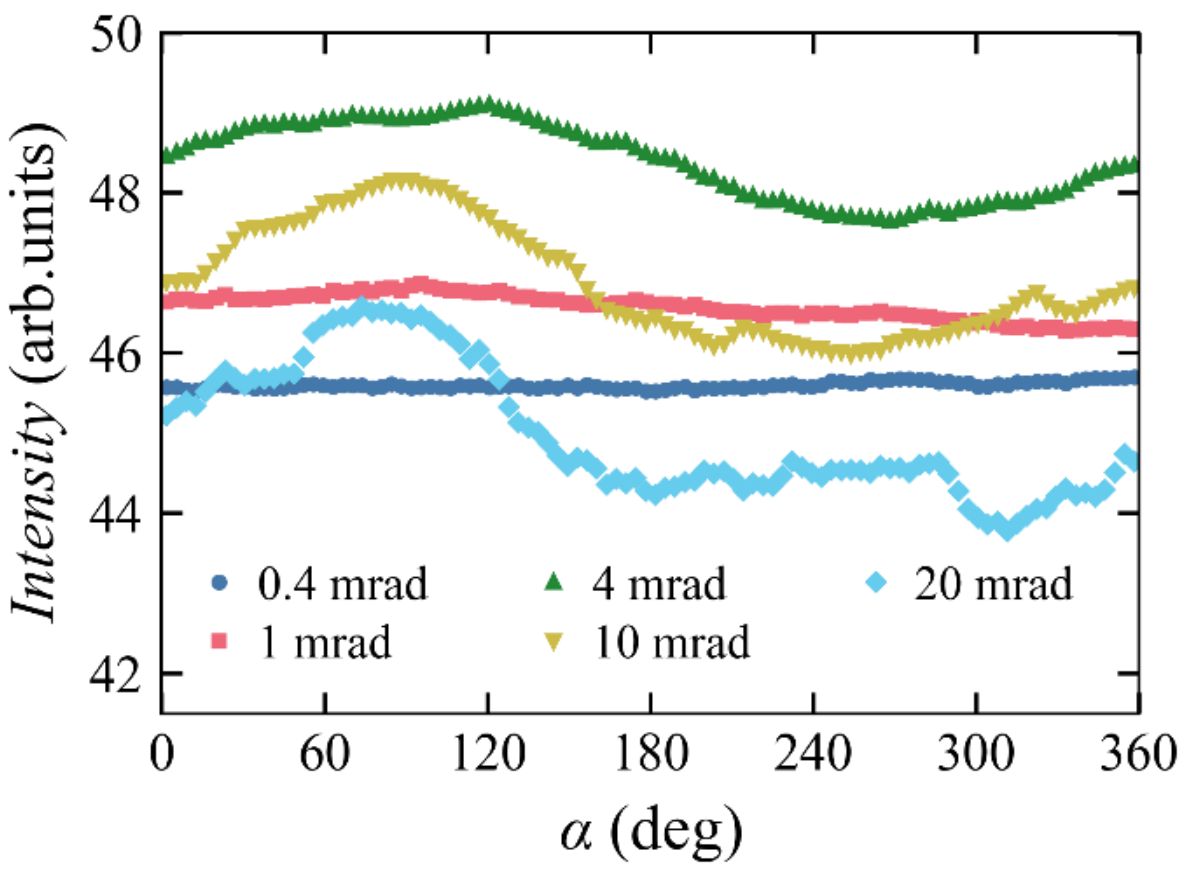

*Figure 8: Transmitted intensity measured in RM734 at 125 °C and f = 10 Hz between crossed polarizers as a function of the phase $\alpha$ of the oscillatory shear for different $\Delta\phi$ oscillation amplitudes.*

Surprisingly, our observations show highly augmented $I_{1f}$ signals of DIO at the N-SmZ$_A$ and SmZ$_A$-N$_F$ transitions. These indicate much larger coupling between the flow and director rotation (enhanced flow alignment) at these transitions. As the SmZ$_A$-N$_F$ transition is approaching, the periodicity of the smectic layers with opposite polarization directions in subsequent layers increases critically[23]. This softens the director fluctuations leading to the enhanced flow alignment. The enhancement at the N-SmZ$_A$ also suggests that the N phase already has N$_F$-like aggregations that critically increase near the transition. The existence of such domains may also explain why, although the measured current is small, the values are out of the measurement error (see Figure 2) in the N phase.

Finally, we would like to point out that both the piezoelectric signals and the second harmonic generation (SHG) nonlinear optical signals rely on the lack of inversion symmetry and the units of SHG and piezo signals are the same. Interestingly, the nonlinear optical (NLO) coefficients $d_{33}$ of RM734 were reported to be in the $pm/\mathrm{V}$ [32,33] range, while for DIO about one order of magnitude smaller values were found [33,34]. Although the m/V and C/N units are the same, the piezoelectric values obtained by us are almost 6 orders of magnitude larger than the SHG coefficients. This discrepancy again can be explained the strong frequency dependences (10 Hz in our and > 400 THz in NLO measurements). Additionally, the underlying physical mechanisms are different as NLO properties originate in the molecular electronic system.

## V. Conclusions

In this work, we have carried out periodic shear-induced electric current measurements on the two archetypic ferroelectric nematic compounds, RM734 and DIO. From temperature, frequency and shear angle dependent results of the first and second harmonic current signals together with temperature and oscillatory rheology measurements we were able to deduce the piezoelectric coupling constants. These values are similar to both materials and were compared to results of previous converse piezoelectric measurements and nonlinear optical (second harmonic generation) coefficients. We also proposed a physical mechanism related to flow alignment of the ferroelectric polarization leading to the direct piezoelectric signals. Our results indicate that about two orders of magnitude increase of the shear-induced electric current based on direct piezoelectric effect of $N_F$ materials can in principle be achieved if the ferroelectric polarization could be aligned uniformly in macroscopic (> 1 cm) size ranges.

## VI. Acknowledgement

This work was financially supported by the Hungarian National Research, Development, and Innovation Office under grants NKFIH FK142643, 2023-1.2.1-ERA_NET-2023-00008 (EIG CONCERT-Japan "FerroFluid") and by the US National Science Foundation under grant DMR-2210083. The work was also supported by a bilateral mobility project (NKM2022-15/2023) between the Hungarian Academy of Sciences and the Japan Society for the Promotion of Science.